\documentclass[%
  reprint,
 superscriptaddress,
bibnotes, 
 amsmath,amssymb,
 aps,
pra,
]{revtex4-2}
\usepackage{lmodern} 
\usepackage{orcidlink}
\usepackage{graphicx}
\usepackage{physics}
\usepackage{xspace}
\newcommand{\ps}{phase space\xspace}

\newcommand{\state}{{{\varrho}}}
\newcommand{\cf}{{\cal S}}

\newcommand{\fdisc}{\emph{faithfully discriminating}\xspace}
\usepackage{hyperref}
\usepackage{color}
\usepackage{soul}
\usepackage[normalem]{ulem}
\newcommand{\VEC}[1]{{\mbox{\boldmath${#1}$}}}

\makeatletter
\newsavebox{\@brx}
\newcommand{\llangle}[1][]{\savebox{\@brx}{\(\m@th{#1\langle}\)}%
  \mathopen{\copy\@brx\kern-0.5\wd\@brx\usebox{\@brx}}}
\newcommand{\rrangle}[1][]{\savebox{\@brx}{\(\m@th{#1\rangle}\)}%
  \mathclose{\copy\@brx\kern-0.5\wd\@brx\usebox{\@brx}}}
\makeatother

\usepackage{forloop,ifthen} 

\usepackage{xifthen}
\newcommand{\refAppendix}[6]{#1
  \ifthenelse{\isempty{#2}}%
    {}
    {\protect\cite{#2}}
    #3\protect\ref{#4}#5#6\xspace
}
\usepackage{tcolorbox}
\definecolor{skyblue}{rgb}{0.53, 0.81, 0.92}

\begin{document}

\title{A Sensitive Nonclassicality Certification Functional for Continuous-Variable Systems}

\author{Ole Steuernagel\orcidlink{0000-0001-6089-7022}}
\email{Ole.Steuernagel@gmail.com}
\affiliation{Institute of Photonics Technologies, National Tsing Hua University, Hsinchu 30013, Taiwan}

\author{Ray-Kuang Lee\orcidlink{0000-0002-7171-7274}}
\affiliation{Institute of Photonics Technologies, National Tsing Hua University, Hsinchu 30013, Taiwan}
\affiliation{Department of Physics, National Tsing Hua University, Hsinchu 30013, Taiwan}
\affiliation{Physics Division, National Center for Theoretical Sciences, Taipei 10617, Taiwan}
\affiliation{Center for Quantum Science and Technology, Hsinchu, 30013, Taiwan}
 
\date{\today}
\begin{abstract}
  If the \ps-based Glauber-Sudarshan distribution, $P_{\state}$, has negative values the quantum
  state,~$\state$, it describes is nonclassical. Due to $P$'s singular behavior, this simple
  criterion is impractical to use.~Recent work has presented a general, sensitive, and
  noise-tolerant certification functional,~$\xi[P]$, for the detection of non\-classical behavior of
  quantum states ${\state}$. There, it was shown that when this functional takes on negative values
  somewhere in \ps,~$\xi[P](x,p) < 0$, this is sufficient to certify the nonclassicality of a
  state. Here we give examples where this certification fails. We investigate states which are known
  to be nonclassical but the certification function is nonnegative, $\xi(x,p) \geq 0$, everywhere in
  \ps.~We generalize $\xi$, giving it an appealing form,~$\cf$, which allows for slight improvements
  in certification, but~$\cf$ also fails for mixed very weakly nonclassical states. More important
  than a slight improvement in sensitivity of~$\cf$ over $\xi$ is that we showed how very sensitive
  $\xi$ and $\cf$ are, and more important still is our simple derivation of~$\cf$ allowing us to
  generalize $\xi$ and $\cf$ to multiple~modes.
  \\
  \emph{Keywords: }{nonclassical state; certification of nonclassical state; phase space
    distributions}
\end{abstract}

\maketitle

\onecolumngrid

\section{Why Nonclassicality Certification Functionals?}

It is desirable to be able to discriminate between nonclassical states on the one hand and classical
states on the other. So far no simple approach that works for theoretical studies and
experimentally reconstructed states alike has been found to perform this discrimination
faithfully. All known approaches to discrimination are either unworkable in the general case, or
they do not work unambiguously, or both~\cite{Froewis_RMP18,Ole_23_Quantumness}.

Here we generalize a recently devised, very sensitive, nonclassicality certification
functional,~$\xi$, introduced in Ref.~\cite{Bohmann_PRL20}. Our generali\-za\-tion provides insight
into the structure of $\xi$ and makes it moderately more \mbox{sensitive} still, yet, the goal of a
universally faithful approach to quantum versus classical discrimination remains elusive.

More importantly, our derivation shows how to extend the approach of~\cite{Bohmann_PRL20} from the
single- to the multi-mode~case.

\subsection*{Nonclassicality Based on the  $P$-Distribution} \label{subsec:_nonclassicality}

In the 1960s it was established that the Glauber-Sudar\-shan
\ps~distribution,~$P$~\cite{Sudarshan__PRL63,Glauber__PR63,Cahill_PR69b,Mandel__PS86,
  Schleich_01,Scully_Zubairy__Book01,Leonhardt_PQE95}, can be used for the very definition of
nonclassical behavior~\cite{Cahill_PR69b,Mandel__PS86,Bohmann_PRL20}.  If a
state,~$\state = \tfrac{1}{2\pi}\iint dx dp P_{\state}(x,p) |\tfrac{x+ip}{\sqrt{2}}\rangle \langle
\tfrac{x+ip}{\sqrt{2}} |$, has a distribution $P_{\state}(x,p)$ with negative values somewhere at
position~$x$ and momentum~$p$ in (cartesian) \ps~({Instead} %
of using complex coherent amplitudes to parameterize \ps~we follow the convention for
(quantum-mechanical) \ps~of Ref.~\cite{Leonhardt_PQE95} using position~$x$ and momentum~$p$.), then
that state is nonclassical.

This, `nonclassicality based on the $P$-distribution', is the approach we adopt here~({Since} the
$P$-distribution can be very singular we recommend that a reader might want to use the equivalent
definition: a state is nonclassical if any \ps~distribution~${\cal P}(1-s)$ of
Equation~(\ref{eq:PDist_R_S}) has negative values.).


So, the Glauber-Sudarshan distribution~$P$ by itself, in principle, successfully discriminates
between classical and nonclassical states, but practically speaking, it does not solve our discrimination
problem. Both, analytically~\cite{Cahill_PR69b} and numerically~\cite{Lvovsky_Raymer__RMP09} $P$ is
hard to determine. Typically, it is so singular that it is extremely difficult to analyze or even
construct $P$~\cite{Wuensche_JOBQSO04,Brewster_Franson__JMP18}.

Useful alternatives to Glauber-Sudarshan's $P$ distribution are Wigner's, $W$~\cite{Wigner_PR32},
and Husimi's, $Q$~\cite{Husimi__PPMSJ40}, distributions. Both are generated through Gaussian
convolutions of the $P$-distribution (see Equation~(\ref{eq:PDist_R_S}) in
Section~\ref{subsec:_smeared_P}), and therefore more well-behaved functions than $P$ is.  But this
convolutional smearing washes out details effectively destroying information about the negative
values of $P$ such that neither~$W$ nor~$Q$ can serve as a function to certify that a state is
nonclassical. $Q$ has lost the ability to develop negative values and also in the case of~$W$ too
much Gaussian-convolution-induced smearing has happened: $W(x,p) <0$ is sufficient to certify that a
state is nonclassical but it is not necessary for a state to have regions with negative values of~$W$ to be
nonclassical~\cite{Bohmann_PRL20,Wuensche_JOBQSO04}.

This situation has led to many different attempts to devise a reliable strategy for certification of
the nonclassical nature of a state
(see~\cite{Luetkenhaus__PRA95,Vogel__PRL00,Diosi__Vogel__PRL00_Comment,Vogel__PRL00_Reply,Asboth__PRL05,Shchukin_Richter_Vogel__PRA05,Miranowicz_PRA10,Park_Zubairy__PNAS17,Bohmann_PRL20,Bohmann_Q20,Chabaud__PRL20,Zhang_Luo__EPJP21,Park_PRR21}
and the many references therein). Similarly, a plethora of measures of
nonclassicality~\cite{Hillery__PRA87,Lee__PRA91,Lee__PRA92,Kenfack_JOB04,Bjoerk__JOBQSO04,Luetkenhaus__PRA95,Manfredi__PRE00,Dodonov_Wuensche__JMO00,Shimizu__PRL02,Marian__PRL02,Duer_Cirac_PRL02,Dodonov_Reno__PLA03,Malbouisson__PS03,Asboth__PRL05,Cavalcanti_Reid__PRL06,Cavalcanti_Reid__PRA08,Boca__PRA09,Lee_Jeong__PRL11,Froewis_Duerr__NJP12,Nimmrichter_Hornberger_PRL13,Yadin_Vedral__PRA16,Tan_Jeong__PRL17,Kwon_Jeong_NJP17,Sekatski__NJP18,Tan_Jeong__PRL20,vanHerstraeten_Cerf__PRA21,Zhang_Luo__EPJP21,Naseri__PRA21}
have been devised.  We list these here because measures are not always distinguished from
certifications~\cite{Luetkenhaus__PRA95,Asboth__PRL05} and also because~$\xi$ lends itself to be
converted to measures with appealing properties~\cite{Ole_23_Quantumness,Chen_26_2modeSqueeExp}.

None of these approaches faithfully discriminate between classical and nonclassical
states: faithful discrimination should always correctly detect and certify that a state is
nonclassical, and it should certify only classical states as classical.  This ability remains
elusive~\cite{Ole_23_Quantumness}.

\begin{figure}[h] \centering
  \includegraphics[width=7.8cm,height=2cm]{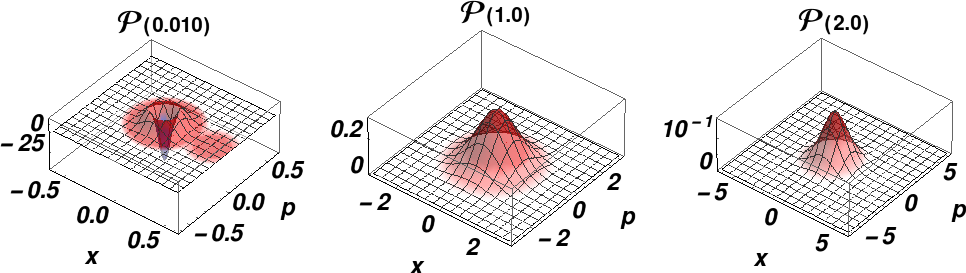}
  \caption{$T$-{parametrized} distributions (see Equation~(\ref{eq:xi_Certification_S_Bohmann_T_parametrization})) %
  for state~(\ref{eq:_BorderCase_state}), with the value
 $w_1 = 0.0122$,  from left to right:
    $P={\cal P}(T=0)$ (approximate), $W={\cal P}(T=1)$ and $Q={\cal P}(T=2)$.}
    \label{Fig:_PWQ_distributions}
\end{figure}

Nonclassicality certification approaches devised so far, typically, mislabel some states that have weak
quantum features as classical~\cite{Lee_Jeong__PRL11,Zhang_Luo__EPJP21}. In the context of \ps-based
approaches, this is often due to the fact that mapping functions from the potentially very singular
$P$-distributions to `better behaved' ones, such as Wigner and Husimi distributions, washes out the
very negative-valued features we intend to detect, see Figure~\ref{Fig:_PWQ_distributions}.

\vspace{-16pt}

The \ps-based certification functional, ${\tilde \xi}(S)$ of Ref.~\cite{Bohmann_PRL20}, or
$\state {\cal L}(\state)$ of Ref.~\cite{Lee_Jeong__PRL11}, use differences of distributions
${\tilde{\cal P}}(S)$, these differences are meant to display areas where ${\tilde \xi}(x,p;S) < 0$
(see Section~\ref{subsec:_smeared_P} below), if and only if $P$ describes a nonclassical state.

To always work, such certification functionals would effectively have to perform Gaussian
deconvolution to retrieve the information held by the $P$-distribution. But Gaussian
de-convolutions are mathematic\-ally and computationally hard to achieve and cannot in general be
performed by using such difference approaches; this can also be illustrated by the fact that
Ref.~\cite{Richter_Vogel__PRL02} shows that in the general case an infinite hierarchy of
certification functionals is required to faithfully discriminate between classical and nonclassical
states. To find a single, general and simple certification functional remains the `holy grail' in
the field of nonclassicality certification and nonclassicality measures. Yet, the approaches of
Ref.~\cite{Bohmann_PRL20} and its general\-ization,~$\cf$, introduced in this work, are surprisingly
versatile and powerful.

In Section~\ref{App:Derive_S} we show that they can faithfully certify all Gaussian states, mixed and
pure. We emphasize that this implies that $\tilde{\xi}(0)$ (based on the Wigner distribution) can
faithfully certify all pure states because Hudson's theorem~\cite{Hudson_RMP74} guarantees that the
Wigner distribution of non-Gaussian pure states has negative regions~\cite{Bohmann_PRL20}.

In Section~\ref{sec:xi_discrimination_fails} we furthermore demonstrate, that, even for very impure
non-Gaussian states, $\xi$ and~$\cf$ give impressive results and that~$\cf$ generalizes~$\xi$.

\section{Smeared Phase Space~Distributions} \label{subsec:_smeared_P}

Smeared \ps~distributions,~${\tilde{\cal P}}(S)$, expressed
in terms of a reference distribution ${\tilde{\cal P}}(R) $ via Gaussian convolutions, are
parameterized by a smearing convolution parameter $S$ and defined as~\cite{Leonhardt_PQE95} %

\begin{flalign}
  {\tilde{\cal P}}(x,p;S) = \frac{1}{{R} -{S} } \int_{-\infty}^{\infty} dx' \int_{-\infty}^{\infty} dp'
  \quad {\tilde{\cal P}}(x',p';R) \times \frac{1}{\pi} \; \exp \left\{ \frac{-1}{{R} -{S} } \left[ ( x -
      x' )^2 + (p - p')^2 \right] \right\}
  \label{eq:PDist_R_S},
\end{flalign}
with allowed parameter values $1 \geq R$ and $R > S$; we see that the Glauber-Sudarshan distribution
$P(x,p) = {\tilde{\cal P}}(x,p; R=1) $ (whilst taking a suitable limit of $S \uparrow 1$) serves as
a starting point of sorts.

Expression~(\ref{eq:PDist_R_S}) connects all $S$-parameterized
\ps~distributions~${\tilde{\cal P}}(S)$.  The~${\tilde{\cal P}}(S)$ are real-valued and with
dropping values of~$S$ increasingly more smeared convolutions of $P$. They include as specially
named cases, the ``exceedingly singular''~\cite{Cahill_PR69b} Glauber-Sudarshan $P$-distribution, as
well as Wigner's distribution $W \equiv {\tilde{\cal P}}(0)$~\cite{Wigner_PR32} and Husimi's
distribution $Q \equiv {\tilde{\cal P}}(-1)$~\cite{Husimi__PPMSJ40}.  Notably, $W$ can have negative
values in \ps~but is a well-behaved function~\cite{Pool__JMP66} and stands out as the only
distribution~${\tilde{\cal P}}(S)$ whose projections exactly give a state's quadratures.  The
smearing in~$Q$ is such that it constitutes the largest possible value of $S$ (giving rise to the
minimal amount of smearing) which is sufficient to guarantee that $Q(x,p) \geq 0$ throughout \ps~for
any nonclassical state $\state$.

\subsection*{Symmetries of $P$-Distribution Nonclassicality}\label{subsec:_PS_symmetries}

Since \ps~displacements and rotations do not change the quantum character of a state, we follow the
quantum optics' community convention that nonclassicality certification functionals should obey the
following symmetries~\cite{Bohmann_PRL20}:

\emph{{Displacement symmetry}%
} in \ps~allows us to always shift states (e.g., to the \ps~origin).

\emph{{Rotation symmetry}} allows us to rotate states (such that they align with coordinate axes).

\section{Bohmann and Agudelo's Phase Space-Based Certification Functional}

The explicit form of the  family of $S$-parameterized certification
functionals devised in Ref.~\cite{Bohmann_PRL20} is
\begin{flalign}
  {\tilde \xi[\tilde{\cal P}]}(S) = {\tilde{\cal P}}(S) - 4 \pi (1-S) \; ({\tilde{\cal P}}(2 S -1))^2, 
  \label{eq:xi_Certification_S_Bohmann}
\end{flalign}
for some details see Appendix~\ref{App:Bohmann_sk}.

For transparency, in subsequent calculations we prefer to employ the \emph{{positive
  $T$-parametrization}}: $T = 1-S$, $T \geq 0$ in Equation~(\ref{eq:PDist_R_S}) whilst dropping the tilde
symbols, used above. Hence,
$\tilde{\cal P}(S) \mapsto {\cal P}(1-S) = {\cal P}(1-(1-T)) = {\cal P}(T) $. With this convention,
the Glauber-Sudarshan distribution is $P \equiv {\cal P}(0)$, the Wigner distribution
$W \equiv {\cal P}(1)$ and Husimi's distribution $Q \equiv {\cal P}(2)$, see
Figure~\ref{Fig:_PWQ_distributions}.

{In expression~(\ref{eq:xi_Certification_S_Bohmann})
${\tilde{\cal P}}(2 S -1) \mapsto {{\cal P}}(1- [2 (1-T) -1]) = {{\cal P}}(2 T) $ which, in
\emph{{positive $T$-parametrization}}, becomes}
\begin{flalign}
    \xi[{\cal P}](T) = 
  {{\cal P}}(T) - 4 \pi \; T \; ({{\cal P}}(2 T))^2.
  \label{eq:xi_Certification_S_Bohmann_T_parametrization}
\end{flalign}

Bohmann and Agudelo showed in Ref.~\cite{Bohmann_PRL20} that all classical single-mode states
obey~$\xi[{\cal P}](x,p;T) \geq 0$. This is to be understood as the statement that the occurrence of
negative values $\xi[{\cal P}](x,p;T) < 0$ somewhere in \ps~$(x,p)$ is \emph{{sufficient}} to show that a
state is nonclassical~\cite{Bohmann_PRL20}.  What is missing for $\xi$ of~\cite{Bohmann_PRL20},
however, is \emph{{necessity}}: to be \fdisc it would be \emph{{necessary}} that whenever a state is
nonclassical we find that $\xi(x,p;T) < 0$ in \ps~somewhere. In
Section~\ref{sec:xi_discrimination_fails} below, we will show that there are states ${\cal P}$ which are
(weakly) nonclassical and yet for these states $\xi[{\cal P}] \geq 0$, across \ps.

\section{Generalizing from $\boldsymbol{\xi}$ to $\boldsymbol{\cf}$}\label{App:Derive_S}

$\xi[{\cal P}](x,p;T)$ is a functional of fixed form independent of context such as the type of system
or the type of state the system is in. $\xi[{\cal P}]$ applies to all states of one-dimensional
continuous-variable systems, irrespective of whether the state is pure or mixed and whether it has
small, large, or even macroscopic~\cite{Leggett__JPCM02} extent, irrespective of context,
hamiltonian, environment or measurement proce\-dures~\cite{Froewis_RMP18}, in this sense $\xi$ is
\emph{{universally}} applicable for all cases and to all states of single-mode systems.

We now generalize~$\xi$ by constructing the certification functional, ${\cf}$, requiring that it
displays negative values certifying a state's nonclassicality, somewhere in phase space. Whereas the
derivation for $\xi$ presented in Ref.~\cite{Bohmann_PRL20} relies on a Chebyshev integral
inequality which only applies to one dimension, our approach is guided by the idea that coherent
states form the dividing line between classical and nonclassical states.  For $\xi$ applied to coherent
states, $\alpha$, this is represented by the \emph{{zero-identity}} $\xi_\alpha(x,p) \equiv 0$. We
thus seek to find a simple way to map coherent {states} %
 onto the vanishing function
$\cf_\alpha(x,p) \equiv 0$.

Consider the $T$-parameterized \ps~distribution of a general Gaussian state
\begin{flalign}
  {\cal P}_g(A,B;T) = %
  {e^{-\frac{(x-x_0)^2}{A}-\frac{(p-p_0)^2}{B}}}/{(\pi \sqrt{A B})} \, .
  \label{eq:GaussState}
\end{flalign}
where, using {{rotation
    symmetry}%
} we assumed alignment with the coordinate axes and using {{displacement symmetry}} we move
its center to the origin ($x_0 = p_0 =0$).

$ {\cal P}_g(A,B;T)$ has respective widths $A$ and $B$; in this representation Heisenberg's
uncertainty principle reads $A B \geq T^2$. If one of these widths, say $A<T$, is below that of the vacuum
state, ${\cal P}_0(T) = {e^{-\frac{x^2}{ T}-\frac{p^2}{ T}}}/{(\pi  T)} $, 
then that state with $A<T$ is squeezed and thus nonclassical~\cite{Wuensche_JOBQSO04}.

Inspired by Ref.~\cite{Bohmann_PRL20}, we form
$ {\cal P}_g(x,p;T+\Delta T) = %
{e^{-\frac{x^2}{A+\Delta T}-\frac{p^2}{B+\Delta T}}}/({\pi
  \sqrt{A+\Delta T} \sqrt{B+\Delta T}})$ from the convolution of $ {\cal P}_g( %
T) $ with ${e^{-\frac{x^2}{\Delta T}-\frac{p^2}{\Delta T}}}/{(\pi \Delta T)} $.
\\ \indent
$ {\cal P}_g(T+\Delta T) $ is wider and its maximum lower than that of $ {\cal P}_g( T) $. 
\\ \indent
To compensate for this increase in width we exponentiate $ {\cal P}_g(T+\Delta T)$ by
$(1 + \Delta T/T)$ which maps the width of coherent states back to $T$.  Subsequent multiplication
by the factor
$(T+\Delta T)^{\frac{T+\Delta T}{T}} ((A+\Delta T) (B+\Delta T))^{-\frac{T+\Delta T}{2 T}} $
increases the height of a coherent state represented by $ {\cal P}_g(T+\Delta T)^{(1 + \Delta T/T)}$
back to that of a coherent state represented by $ {\cal P}_g(T)$. Forming the difference gives us

\begin{equation} 
\begin{array}{l}
  \!\!\! {\cf}[{\cal P}_g(A,B;T)](x,p;T,\Delta T)\\
  = \frac{e^{-\frac{x^2}{A}-\frac{p^2}{B}}}{\pi
  \sqrt{A B}}-\frac{e^{{ \left(-\frac{x^2(T+\Delta T)}{(A+\Delta T)T}-\frac{p^2(T+\Delta
  T)}{(B+\Delta T)T}\right)}}}{\pi T}  \times (T+\Delta T)^{\frac{T+\Delta T}{T}} ((A+\Delta T) (B+\Delta T))^{-\frac{T+\Delta T}{2 T}}.
\end{array}
\label{eq:deriveS}
\end{equation}

Inserting a coherent state ($A=B=T$), yields ${\cf}[{\cal P}_g(T,T;T)](T,\Delta T) \equiv 0$, as
desired.  \\\noindent Re-expressed in terms of $T$-parameterized {{general
    states}}~${\cal P}$, expression~(\ref{eq:deriveS}) gives us the certification functional
\begin{flalign}
  {\cf}[{\cal P}](x,p;T, {\Delta T}) 
  = & \;
  {\cal P}\left(x,p;T\right)-\frac{{\pi}^{\frac{ {\Delta T}}{T}}}{T}
  \left[ \left( T+{\Delta T} \right) {\cal P}\left(x,p; T+{\Delta T} \right)\right]{}^{\frac{T+ {\Delta T}}{T}}
  \label{eq:SmoothingMeasure} \\
  = & \; {\cal P}\left(x,p;T\right) - N_{0} \left(T\right) \left( \frac{{\cal P}\left(x,p; T+{\Delta
          T} \right)} { N_{0} \left( T+{\Delta T} \right)} \right){}^{\frac{ T+{\Delta T} }{T}}
   \text{, \quad where } N_0(T) = \tfrac{1}{\pi T}.
  \label{eq:SmoothingMeasure2}
\end{flalign}

We emphasize that $ {\cf}[{\cal P}]$ should only be applied to \emph{nonnegative states}, namely,
states which obey ${\cal P}(x,p;T+\Delta T) \geq 0 $, since non-integer exponents $(T+\Delta T)/T$
can generate complex values if a state has negative values somewhere in \ps.

This restriction to nonnegative states is an acceptable price to pay for two reasons: Firstly,
\emph{negative states}, namely those for which ${\cal P}(x,p;T+\Delta T) < 0 $ somewhere in \ps, are
already certified to be non-classical by displaying their negative values. Secondly, for the special
case of $T=\Delta T$ we have that ${\cf}[{\cal P}](T,T) = \xi(T)$ with the associated exponent
$(T+\Delta T)/T =2$ which allows us to form the difference in Expression~(\ref{eq:SmoothingMeasure})
as desired, also for negative states, and, as we show in Section~\ref{sec:xi_discrimination_fails},
$\xi(T)$ tends to work nearly as well as ${\cf}[{\cal P}](T, \Delta T)$.

When $ {\Delta T} = T$, $ {\cf}(T,T) = \xi(T)$, see
Equation~(\ref{eq:xi_Certification_S_Bohmann_T_parametrization}).  Thus, $ {\cf}(T, {\Delta T})$
generalizes $\xi(T)$, see Section~\ref{sec:xi_discrimination_fails}.

\subsection*{Gaussian States}

A particular strength of the certification function $\xi$ of Ref.~\cite{Bohmann_PRL20} is its
ability, without fail, to detect the nonclassical~\cite{Wuensche_JOBQSO04} squeezing of
fluctuations below the vacuum level in \emph{{all Gaussian states}}; irrespective of whether the
states are pure or impure states. This follows from a slight modification of the arguments given
when deriving Equation~(\ref{eq:SmoothingMeasure}):

When deriving Equation~(\ref{eq:SmoothingMeasure}), we were guided by the principle that we aim to
map pure coherent states, the states that define the boundary between classical and quantum states,
onto the identically vanishing function ${\cf} \equiv 0$. It is plausible then that ${\cf}$ maps
nonclassical Gaussian states to functions which in parts of \ps~fall `below' this zero-function,
forming negative regions, whereas ${\cf}$ maps classical states to functions greater than zero
everywhere in \ps, see~Section~\ref{sec:xi_sees_classicalStates}.

\begin{figure}[h] \centering
  \includegraphics[width=7.4cm,height=4.25cm]{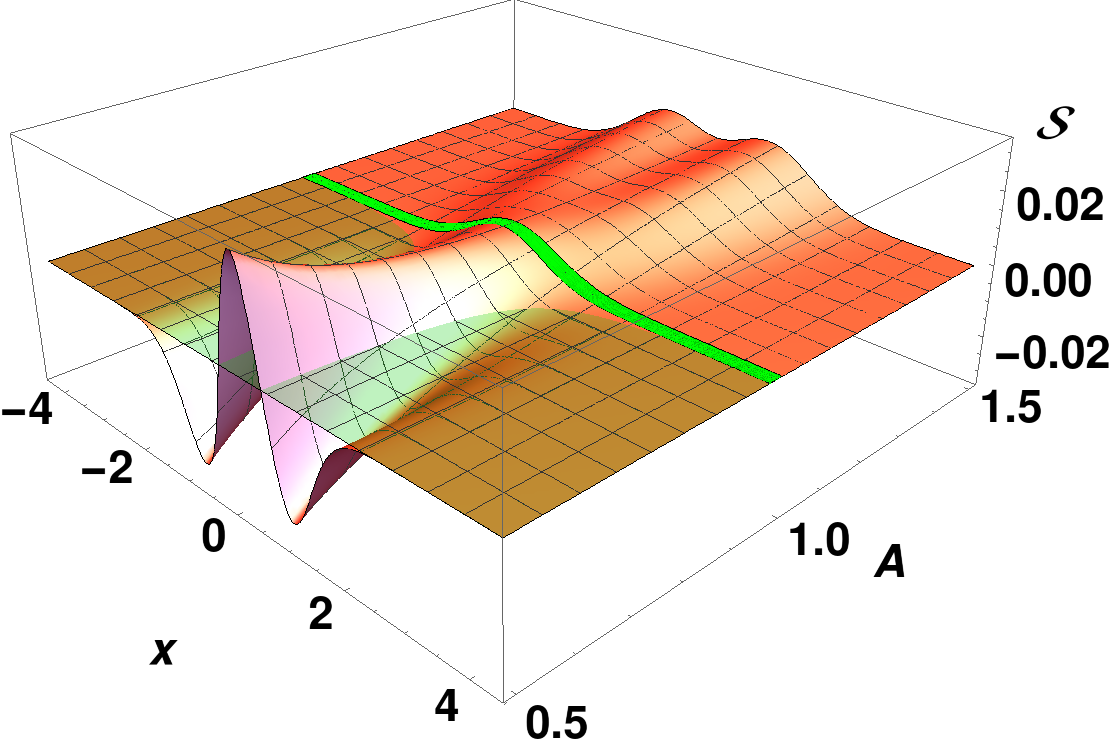}
  \caption{\mbox{${\cf}[{\cal P}_g(A,B=2;T=1)](x,p=0;T=1,\Delta T = \tfrac{3}{4})$} based on the Wigner
    distribution $(T=1)$ of the impure squeezed states ${\cal P}_g(A,B=2;T=1)$ as a function of the
    squeezed coordinate $x$ with varied $x$-squeezing width $A$ and fixed $p$-coordinate
    anti-squeezing width $B=2$. The green line at $T=1$ divides values of $A<T$ where the
    nonclassical character of the state is always certified by the occurrence of negative values of
    ${\cf}$, whilst for $A \geq T$, ${\cf} \geq 0$, just as desired.}
    \label{Fig:_S_gauss_perfect}
  \end{figure}

Let us first consider nonclassical Gaussian states:

For Gaussians which (in their squeezed coordinate) decay faster than vacuum states, the spreading
when mapping $ {\cal P}\left(T\right)$ to ${\cal P}\left(T+{\Delta T} \right)$ is not fully
compensated for by the narrowing due to the exponentiation by $E = {{T+ {\Delta T}}/{T}} $:
therefore, in the squeezed direction, the $ {\cal P}\left(T\right)$-term in
Equation~(\ref{eq:GaussState}) decays faster than the $ {\cal P}\left(T + {\Delta T} \right)$-term
and, upon forming their difference, ${\cf}$ forms regions with negative values on the squeezed
`narrow flanks' of the \ps~distribution~(for illustration see Figure~\ref{Fig:_S_gauss_perfect} and
also Figure~1 of~\cite{Bohmann_PRL20}). ${\cf}[{\cal P}_g(A,B;T)](x,p;T,\Delta T)$ faithfully
certi\-fies suppression of fluctuations to nonclassical \mbox{levels} (here, below the level of
vacuum fluctuations~\cite{Wuensche_JOBQSO04}) for {{all Gaussian states}}.

\vspace{-24pt}

\section{ $\boldsymbol{{\cf}(T, {\Delta T})}$ Generalizes $\boldsymbol{\xi(T)}$}\label{subsec:xi_generalized}

Our derivation also allows us to generalize to the multimode case of 
 $K$ cartesian \ps~modes:
\begin{flalign} 
  {\cf}(T, {\Delta T}) = {\cal P}\left(T\right) - N^K_{0} \left(T\right) \left(
    \tfrac{{\cal P}\left(T+{\Delta T} \right)}
    { N^K_{0} \left( T+{\Delta T} \right)} \right){}^{\tfrac{ T+{\Delta T} }{T}} .
  \label{eq:_S_multimode}
\end{flalign}

This can be confirmed by insertion of the $K$-mode vacuum state, $({\cal P}_{|0\rangle\langle
  0|}(T))^{\otimes K} $ into~(\ref{eq:_S_multimode}) yielding ${\cf}(\VEC{x}, \VEC{p}) \equiv
0$ (where bold fonts,~$\VEC{x}$, describe
$K$-dimensional vectors).

The negative values of Gaussian states form just like in the single-mode case~(\ref{eq:GaussState}),
see Figure~\ref{Fig:_S_gauss_perfect}, such that the results of the single-mode case carry over to the
multi-mode case.  This generalization of the single-mode expression~(\ref{eq:SmoothingMeasure2}) to
the multi-mode expression~(\ref{eq:_S_multimode}) is the main result of this work.

\subsection{$\xi(T)$ and $ {\cf}$ Can always Identify Classical States}\label{sec:xi_sees_classicalStates}

$ {\cf}$, just like $\xi$, without fail, certifies classical states (incoherent mixtures of
Glauber coherent states) as classical, since the nonlinear terms in Equation~(\ref{eq:SmoothingMeasure})
for a mixture between states ${\cal P}_1$ and ${\cal P}_2$, parameterized by ($0<w<1)$, form the
positive concave difference function
${\cf}[w {\cal P}_1 + (1-w) {\cal P}_2] - \left(w{\cf}[{\cal P}_1] + (1-w) {\cf}[{\cal
    P}_2]\right) \geq 0 $. This renders mixtures ${\cf}$ more positive than the mixture's
individual contributions, as desired.

We now show this in two steps.

By construction ${\cf} \equiv 0$ for pure coherent states, see Equation~(\ref{eq:deriveS}). Since
classical states are incoherent mixtures of Glauber coherent states, we first establish that a
coherent state contributing to a mixture with below-unit weight $w$ {($0<w<1$) adds a positive
  contribution to ${\cf}[ w {\cal P}_{0}] > 0$:} recall that Equation~(\ref{eq:SmoothingMeasure}) implies
that $ {\cf}[w {\cal P}_{0}](T, {\Delta T}) = w \times$
\mbox{$ \left( {\cal P}_{0}\left(T\right) - {w}^{\frac{{\Delta T} }{T}} N_{0} \left(T\right) \left(
    \frac{{\cal P}_{0}\left(T+{\Delta T} \right)} { N_{0} \left( T+{\Delta T} \right)}
  \right){}^{\!\!\frac{ T+{\Delta T} }{T}} \right) \! > \! w \, {\cf}[{\cal P}_{0}] %
\equiv 0,$} since ${w}^{\frac{{\Delta T} }{T}}<1$.

\subsubsection*{$\xi(T)$ and $ {\cf}$ Are Concave Functionals of the Mixing of
  States}\label{sec:xi_sees_mixednessStates}

In this second step, we now generalize this result by showing, for all states, mixtures render ${\cf}$ more positive than the mixture's individual contributions. This reduces the telltale
negative values of nonclassical, nonnegative states for mixtures.

For nonnegative states ${\cf}$ is concave in the mixing weight $w$ of the mixed state
$w {\cal P}_1 + (1-w) {\cal P}_2$ (with $0<w<1$):
\begin{flalign}
  \Delta {\cf}(T, \Delta T) =\; &{\cf} [w {\cal P}_1 + (1-w) {\cal P}_2] %
   - \Big(w{\cf}[{\cal P}_1] + (1-w) {\cf}[{\cal P}_2]\Big) \\
  \propto & - \left( w {\cal P}'_1
    + (1-w) {\cal P}'_2 \right)^E %
   + \left( w ({\cal P}'_1)^E + (1-w) ({\cal P}'_2)^E \right) > 0 ;
\label{app_eq:concave}
\end{flalign}
here we have used the abbreviations ${\cal P}' = {\cal P}\!\left(T+{\Delta T}\right)$ and
$E = (T+{\Delta T})/T > 1 $ and we assumed that ($ {\cal P}_1 (x,p;T) \geq 0$,
$ {\cal P}_2 (x,p;T) \geq 0$), and ${\cal P}_1 \neq {\cal P}_2$, and that
inequality~(\ref{app_eq:concave}) obeys a Jensen's inequality for concave functions, making
$\Delta {\cf} $ concave under mixing.

These results imply that $\cf < 0 $ (just like $\xi <0$) certifies that a state is nonclassical.

Because multimode classical states can be described by sums of (unentangled) product states these
result carry over to the multimode case of Equation~(\ref{eq:_S_multimode}).

\subsection{$\xi(T)$ Can Be Faithfully Discriminating for Some Classes of States}\label{sec:xi_discriminates}

Additionally to Gaussian states, there are a number of cases when the certification functional
$\xi(T)$ of Ref.~\cite{Bohmann_PRL20} is {{faithfully
    discriminating}%
}.

For example the mixed state $\state = \cos(w)^2 |0\rangle\langle 0| + \sin(w)^2 |1\rangle\langle 1|$,
to first order in $w$, yields
$\xi(T) = {w \; e^{-\frac{p^2+x^2}{T}} \left(p^2+x^2-T^2\right)}/{(\pi T^3)}$: even for vanishing
non\-classical contributions $(w\downarrow 0)$, $\xi$ assumes negative values near the origin,
faithfully detecting the presence of the single photon Fock state $|1\rangle\langle 1|$ in this
mixture. $\xi$ similarly detects the presence of such mixtures if Fock states of order higher than
one are used, and also if those are displaced with respect to the dominant (classical) \mbox{vacuum} state
$\state = |0\rangle\langle 0|$.

For the coherent superposition state $|\psi \rangle = \cos(w) |0\rangle + \sin(w) |1\rangle $, between vacuum
and single-photon Fock state, we get, to second order in $w$, that
$\xi = w^2 \left( p^2-x^2 \right) $ $\times$~exp$[{-\tfrac{p^2+x^2}{T}}]/{(\pi T^3)}$.  Again, if
Fock states of higher order are combined with the dominant (classical) vacuum state and also if they
are displaced with respect to each other, in all these cases, $\xi <0$ in \ps~somewhere,
certi\-fy\-ing the presence of (vanishingly small) nonclassical contributions to the state.

\subsection{$\xi(1)$  Always Detects Non-Gaussian Pure States}

Wigner distributions of non-Gaussian pure states have negative regions~\cite{Hudson_RMP74}. Because
$Q\geq 0$ this implies that, for $T=1,$ $ \xi(1) = W - 4\pi Q^2 < 0$ when $W<0$.

We believe that pure states might exist for which more smeared versions of $\xi(T)$ and
$ {\cf}(T, {\Delta T})$ with $T>1$ fail to certify nonclassicality.

\section{A Case Where First $\boldsymbol{\xi(T)}$ and then $\boldsymbol{{\cf}(T, {\Delta T})}$ Fails
  to Be Discriminating}\label{sec:xi_discrimination_fails}

\begin{figure}[h] \centering
  \includegraphics[width=7.8cm,height=2.8cm]{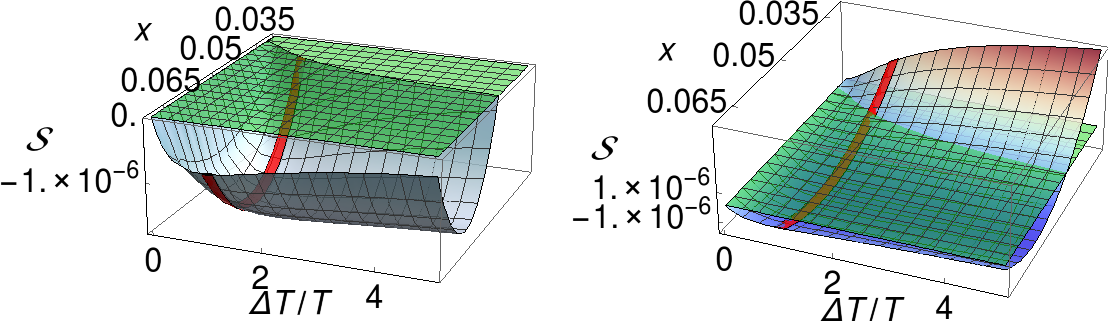}
  \put(-216,81){\rotatebox{0}{{\bf (a) \qquad \quad \bf \textcolor{blue}{$w_1 = 0.0125$} \; (b)}}}\\
  \vspace{0.2cm}
  \includegraphics[width=7.8cm,height=2.8cm]{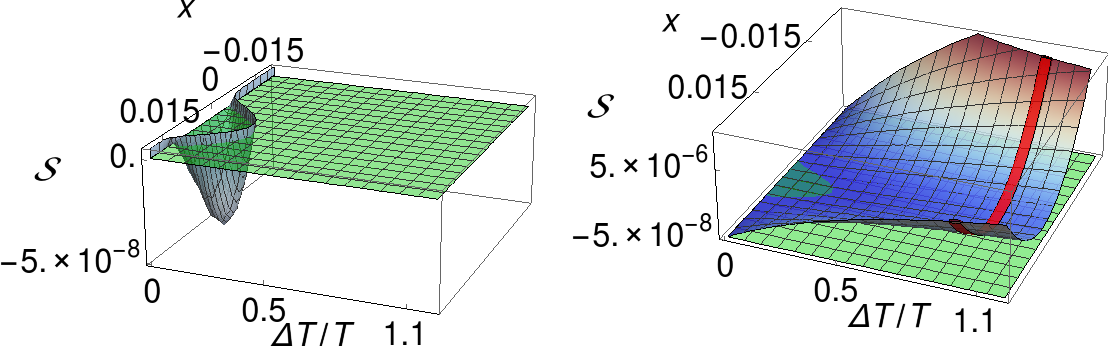}
  \put(-216,79){\rotatebox{0}{{\bf (c) \qquad \quad  \bf \textcolor{blue}{$w_1 = 0.0122$}  \; (d)} }}
  \caption{{Display} %
 of $\xi(T=1)$ (red bands) and ${\cf}(T=1,\Delta T)$ (color sheets) for Wigner's
    distribution, ${\cal P}(T=1)$. We consider the weakly nonclassical
    state~(\ref{eq:_BorderCase_state}): for values of $w_1 = 0.0125$ (top panels (\textbf{a},\textbf{b}))
    both certification functionals $\xi(x,p=0;T=1)$ of {Ref.}~%
 \cite{Bohmann_PRL20}
    and~${\cal S}(x,p=0;T=1,\Delta T)$ of~Equation~(\ref{eq:SmoothingMeasure}) drop to nega\-tive values
    (below green transparent zero-value sheets), certifying the nonclassicality of
    state~(\ref{eq:_BorderCase_state}), see Figure~\ref{Fig:_PWQ_distributions}. But for the value
    $w_1 = 0.0122$ (bottom panels (\textbf{c},\textbf{d})), whereas the certification functional
    ${\cf}(x,0;T=1,\Delta T)$ still {shows} %
 negative values, $\xi$ fails to certify that this state,
    depicted in Figure~\ref{Fig:_PWQ_distributions}, is nonclassical (red band does not dip below
    green zero-sheet). For slightly smaller values of $w_1 < 0.0122$ (not shown) ${\cf}$ fails as
    well. In short, ${\cf}$ can be more discriminating than~$\xi$ but only very moderately so.}
    \label{Fig:best_discrimination_dT}
\end{figure}

In order to investigate a scenario where $\xi(T)$ fails to be discriminating, we consider a state
with a mixed positive background distribution consisting of a vacuum state, $W_0(x,p)$, overlapping
with a coherent state, $W_0(x - \tfrac{19}{40},p)$, and a (nonclassical) single-photon Fock state,
$W_1$, which carries a small weight $w_1 \approx 1$\%, namely:
\begin{flalign}
\!\!  W(w_1&,s)  = w_1 W_1(x,p) + (1 - w_1) \Big( W_0 ( x , p) \cos^2(s) + \sin^2(s) W_0 ( x - \tfrac{19}{40}, p) \Big) , \!
\label{eq:_BorderCase_state}
\end{flalign}
with $s=\tfrac{\pi}{20}$. This state is nonclassical, see
Figure~\ref{Fig:_PWQ_distributions}. Figure~\ref{Fig:best_discrimination_dT} shows that for small values
of $\Delta T$ the certification functional ${\cf}(T, \Delta T)$~(\ref{eq:SmoothingMeasure}) can be
slightly more discriminating than ${\cf}(T,T)= \xi(T)$ of Ref.~\cite{Bohmann_PRL20}, but both fail
when the single-photon state's weight $w_1$ drops further, below 0.0122. This is due to
inequality~(\ref{app_eq:concave}).

The example state~(\ref{eq:_BorderCase_state}) appears contrived and
Figure~\ref{Fig:best_discrimination_dT} reports only a very modest increase in sensitivity of our
certification functional ${\cf}$ over $\xi$. Instead, of taking this point of view we advise
to adopt the following perspective:

Figure~\ref{Fig:best_discrimination_dT} firstly confirms that ${\cf}(T,\Delta T)$ really differs from
$\xi(T)$, hence, ${\cf}$ truly generalizes $\xi$ of Ref.~\cite{Bohmann_PRL20}.

Secondly, it
illustrates how very discriminating between classical and nonclassical states ${\cf}$ and $\xi$ really
are, namely, a roughly 1.2~\!\% nonclassical contribution to state~(\ref{eq:_BorderCase_state}) is
detected; also see Section~\ref{sec:xi_discriminates}.

And, thirdly, when we apply ${\cf}$, typically
little is gained by using $\Delta T \neq T$ in ${\cf}(T,\Delta T)$ instead of sticking with
$\Delta T = T$ such that ${\cf}(T,T) =\xi (T)$. Therefore, sticking with $\xi$ and its
multimode generalization $\cf(T,T)$ of~Equation~(\ref{eq:_S_multimode}) is typically good enough.

\section{Conclusions}\label{sec:Conclusion}

We have generalized the nonclassicality certification functional of Ref.~\cite{Bohmann_PRL20}. We
have proved that the nonclassicality functionals~(\ref{eq:SmoothingMeasure2}) always identify
classical states correctly, but for weakly nonclassical states they can fail, incorrectly
mislabeling these as classical. We proved that these nonclassicality certification functionals
always identify mixtures of positive-valued states as the functionals are {{concave}} in the mixing
probabilities of the states' constituents, see Equation~(\ref{app_eq:concave}). Our generalized
certification
functionals' transparent form hopefully opens up the door for further improvements eventually
resulting in a general and useful discriminating approach to nonclassicality certification. Its
transparent form has allowed us to generalize to multimode settings, see
Equation~(\ref{eq:_S_multimode}).

		\section*{Author Contributions}
                
                {O.S. and R.-K.L.: Conceptualization, methodology, writing—reviewing and editing; O.S.: Software, writing—original draft preparation, visualization. Both authors have read and agreed to the published version of the manuscript.}

		\section*{Funding}
		
				This work is partially supported by the National Science and Technology Council of Taiwan (Nos 112-2123-M-007-001, 112-2119-M-008-007, 114-2112-M-007-044-MY3), Office of Naval Research Global, the International Technology Center Indo-Pacific (ITC IPAC) and Army Research Office, under Contract No. FA5209-21-P-0158, and the collaborative research program of the Institute for Cosmic Ray Research (ICRR) at the University of Tokyo.
 
		\section*{Institutional Review Board Statement}
		
                Not applicable.
                
		\section*{Informed Consent Statement}

                Not applicable.

		\section*{Data Availability Statement}
		
                Not applicable.
                
		\section*{Acknowledgments}
		
                We feel indebted to an anonymous Editor for pointing out that $\xi$ of
                Ref.~\cite{Bohmann_PRL20} is not always {{faithfully discriminating}}, triggering
                the present investigations. We thank the production staff of this journal for their
                extremely valuable feedback on this manuscript.

		\section*{Conflicts of Interest}

                The authors declare no conflict of interest.

		\section*{Use of AI and AI-Assisted Technologies}

                No AI tools were utilized for this paper.

\begin{appendix}
\section{Smeared Unbalanced Certification Functionals~$\boldsymbol{{\tilde \xi}(S,k)}$}\label{App:Bohmann_sk}

In Ref.~\cite{Bohmann_PRL20} the two-parameter families of nonclassicality
certification functions 
\begin{flalign}
  {\tilde \xi}(S,k) = {\tilde {\cal P}}(S) - \frac{\pi(1-S)}{k(1-k)} \; {\tilde {\cal P}}(S_k) \; {\tilde {\cal P}}(S_{1-k})
  \label{eq:xi_Certification_Sk_Bohmann_appendix} ,
\end{flalign}
with $S_k = 1 - \frac{1-S}{k}$, were introduced. Here the parameter $S\leq 1$ selects the reference
distribution~${\tilde {\cal P}}(x,p;S)$, see Equation~(\ref{eq:PDist_R_S}).
Expression~(\ref{eq:xi_Certification_Sk_Bohmann_appendix}) can be useful for some field correlation
measurements~\cite{Bohmann_PRL20}.

The parameter $k \in (0,1)$ linearly interpolates between different exponents in the distributions
${\tilde {\cal P}}(S_k)$ \mbox{${\tilde {\cal P}}(S_{1-k})$}
$ = {\tilde {\cal P}}(S_{1-k}) {\tilde {\cal P}}(S_{k})$ which are smeared with respect
to~${\tilde {\cal P}}(S)$. Note that, because of the symmetry in $k$ with respect to the
midpoint~$k=\frac{1}{2}$, effectively $k \in (0, 0.5]$.

Also, for vanishing $k$ the limit is $\lim_{k\downarrow 0} S_k= - \infty$ and
$\lim_{k\downarrow 0} S_{1-k} = S$ and we get, according to Equation~(\ref{eq:PDist_R_S}), an infinitely
spread-out distribution~${\tilde {\cal P}}(-\infty)$ (giving us an essentially constant background,
which vanishes with increasing smearing values $|S|$) multiplied with~${\tilde {\cal P}}(S)$: in
short, for vanishing $k$, in \ps~regions where it assumes negative values,
expression~(\ref{eq:xi_Certification_Sk_Bohmann_appendix}) rises towards zero.

As a function of $k$, negative minimum values ${\tilde \xi}_-(S_k)$ with the grea\-test magnitude,
occur at the balanced mid-point, $k=\frac{1}{2}$, providing the grea\-test contrast when applying
$\tilde \xi$~\cite{Ole_23_Quantumness}. This is why here, without loss of generality, we confine our
discussion of $\xi$ to the form of Equation~(\ref{eq:xi_Certification_S_Bohmann}) only.

\end{appendix}

\end{document}